\documentclass[11pt]{article} 
\makeatletter\renewcommand{\section}{\@startsection
	{section}{1}{\z@}{-3.5ex plus -1ex minus
		-.2ex}{2.3ex plus .2ex}{\bf }}
\makeatletter\renewcommand{\subsection}{\@startsection{subsection}{2}{\z@}{-3.25ex
		plus -1ex minus
		-.2ex}{1.5ex plus .2ex}{\it }}
\makeatletter\renewcommand{\subsubsection}{\@startsection{subsubsection}{3}{-2.45ex}{-3.25ex
		plus -1ex minus -.2ex}{1.5ex plus .2ex}{\it }}

\makeatletter \@addtoreset{equation}{section}
\usepackage[table,xcdraw]{xcolor}
\renewenvironment{thebibliography}[1]
{\baselineskip=16pt plus 2pt minus 1pt
	\section*{\large\refname
		\@mkboth{\MakeUppercase\refname}{\MakeUppercase\refname}}%
	\list{\@biblabel{\@arabic\c@enumiv}}%
	{\settowidth\labelwidth{\@biblabel{#1}}%
		\leftmargin\labelwidth
		\advance\leftmargin\labelsep
		\@openbib@code
		\usecounter{enumiv}%
		\let\p@enumiv\@empty
		\renewcommand\theenumiv{\@arabic\c@enumiv}}%
	\sloppy
	\clubpenalty4000
	\@clubpenalty \clubpenalty
	\widowpenalty4000%
	\sfcode`\.\@m}

\let\fn\footnote
\renewcommand{\footnote}[1]{\linespread{1.1}\fn{#1}\linespread{1.29}}

\usepackage{multirow}
\usepackage{epsfig} 
\usepackage{verbatim}
\usepackage{graphicx}
\usepackage{tikz}
\usetikzlibrary{trees,er,snakes,shapes,mindmap}
\usepackage{hyperref}
\usepackage{easyfig}
\usepackage{bbm}
\usepackage{cite}
\usepackage{slashed}
\usepackage{physics}
\usepackage{amsmath}
\usepackage{amsfonts}
\usepackage{amssymb}
\usepackage{braket}
\usepackage{mathrsfs}
\usepackage{bm}
\usepackage[section]{placeins}
\usepackage{placeins}
\usepackage{mathtools}
\usepackage[utf8x]{inputenc}
\usepackage{booktabs}
\usepackage{caption}
\usepackage{tensor}
\usepackage{dsfont}

\def\tyng(#1){\hbox{\tiny$\yng(#1)$}}

\let\Oldsection\section
\renewcommand{\section}{\FloatBarrier\Oldsection}

\let\Oldsubsection\subsection
\renewcommand{\subsection}{\FloatBarrier\Oldsubsection}

\let\Oldsubsubsection\subsubsection
\renewcommand{\subsubsection}{\FloatBarrier\Oldsubsubsection}

\allowdisplaybreaks
\newcommand{\be}{\begin{equation}}
\newcommand{\ee}{\end{equation}}
\newcommand{\beq}{\begin{equation}}
\newcommand{\eeq}{\end{equation}}
\newcommand{\bea}{\begin{array}}
	\newcommand{\ea}{\end{array}}
\newcommand{\beqa}{\begin{eqnarray}}
\newcommand{\eeqa}{\end{eqnarray}}
\newcommand{\beqar}{\begin{eqnarray}}
\newcommand{\eeqar}{\end{eqnarray}}
\newcommand{\nn}{\nonumber}
\begin{document}
%%%%%%%%%%%%%%%%%%%%%%%%%%%%%%%%%%%%%%%%%%%%%%%
%%%%%%%%%%%%%%%%%%%%%%%%%%%%%%%%%%%%%%%%%%%%%%%
%\fontfamily{pnb}\fontsize{12pt}{16pt}\selectfont
%\fontfamily{pzc}\fontsize{14pt}{16pt}\selectfont
%\fontfamily{pbk}\fontsize{12pt}{16pt}\selectfont
%\fontfamily{cmr}\fontsize{11pt}{15pt}\selectfont
\fontfamily{bch}\fontsize{11pt}{15pt}\selectfont
%\fontfamily{phv}\fontshape{ro}\fontsize{11pt}{14pt}\selectfont
%\fontfamily{ptm}\fontseries{m}\fontshape{r}\fontsize{12pt}{16pt}\selectfont
%\fontfamily{pnc}\fontseries{m}\fontshape{r}\fontsize{11pt}{15pt}\selectfont
%\fontfamily{ppl}\fontseries{m}\fontshape{r}\fontsize{11pt}{15pt}\selectfont
%\usefont{T1}{phv}{m}{it}
%%%%%%%%%%%%%%%%%%%%%%%%%%%%%%%%%%%%%%%%%%%%%%%
%%%%%%%%%%%%%%%%%%%%%%%%%%%%%%%%%%%%%%%%%%%%%%%
	\begin{titlepage}
		\begin{flushright}
			%{\bf Draft 9} { \bf  \today} 
		\end{flushright}
		
		%\vskip 1 em
		
		\begin{center}
{\Large \bf  Extended Dynamical Symmetries of Landau Levels in Higher Dimensions}\\
~\\
%{\Large \bf }
			
			%\centerline{{\Large \bf }}
			
			\vskip 1em
			
			\centerline{ $ \text{\large{\bf{S. K\"{u}rk\c{c}\"{u}o\v{g}lu}}}^{a} \, \, $, $ \text{\large{\bf{G.\"{U}nal }}}^{b} \, \, $, $ \text{\large{\bf{İ. Yurdu\c{s}en}}}^{c} $}
			
			\vskip 0.5cm
			\centerline{\sl $^a$ Middle East Technical University, Department of Physics,}
			\centerline{\sl Dumlupınar Boulevard, 06800, Ankara, Turkey}
			\vskip 1em
				\centerline{\sl $^b$ İzmir Institute of Technology, Department of Physics, IZTECH, 35430,}
			\centerline{\sl İzmir, Turkey }
			\vskip 1em
			\centerline{\sl $^c$ Hacettepe University, Department of Mathematics, 06800, Beytepe,}
			\centerline{\sl Ankara, Turkey}
			\vskip 1em
 \vskip .26cm
\begin{tabular}{r l}
E-mail:&\!\!\!{\fontfamily{cmtt}\fontsize{11pt}{15pt}\selectfont kseckin@metu.edu.tr}\\
&\!\!\!{\fontfamily{cmtt}\fontsize{11pt}{15pt}\selectfont gonulunal@iyte.edu.tr }\\
&\!\!\!{\fontfamily{cmtt}\fontsize{11pt}{15pt}\selectfont yurdusen@hacettepe.edu.tr}
\end{tabular}
			
		\end{center}
		
		\vskip 1 em
		
		\begin{quote}
			\begin{center}
				{\bf Abstract}
\end{center}
			
\vskip 1em

Continuum models for time-reversal (TR) invariant topological insulators (TIs) in $d \geq 3$ dimensions are provided by harmonic oscillators coupled to certain $SO(d)$ gauge fields. These models are equivalent to the presence of spin-orbit (SO) interaction in the oscillator Hamiltonians at a critical coupling strength (equivalent to the harmonic oscillator frequency) and leads to flat Landau Level (LL) spectra and therefore to infinite degeneracy of either the  
positive or the negative helicity states depending on the sign of the SO coupling. Generalizing the results of \cite{Haaker} to $d \geq 4$, we construct vector operators commuting with these Hamiltonians and show that $SO(d,2)$ emerges as the non-compact extended dynamical symmetry. Focusing on the model in four dimensions, we demonstrate that the infinite degeneracy of the flat spectra can be fully explained in terms of the discrete unitary representations of $SO(4,2)$, i.e. the {\it doubletons}. The degeneracy in the opposite helicity branch is finite, but can still be explained exploiting the complex conjugate {\it doubleton} representations. Subsequently, the analysis is generalized to $d$ dimensions, distinguishing the cases of odd and even $d$. We also determine the spectrum generating algebra in these models and briefly comment on the algebraic organization of the LL states w.r.t to an underlying ``deformed" AdS geometry as well as on the organization of the surface states under open boundary conditions in view of our results.

%We study the four-dimensional harmonic oscillator with an $SO(4)$ gauge field background, equivalent to the presence of  spin-orbit (SO) term in the Hamiltonian at a critical coupling strength (equivalent to the Harmonic oscillator frequency) leading to an infinite-fold degeneracy in the branch in which orbital and spin angular momentum are aligned (i.e. in the positive SO branch). Generalizing the ideas of  \cite{}, we construct the Runge-Lenz type vectors commuting with this Hamiltonian and show that the latter possesses $SO(4,2)$ as a non-compact dynamical symmetry group. We show that the infinite degeneracy of the Hamiltonian in the positive SO branch can be fully explained in terms of the discrete unitary representations of $SO(4,2)$, which are also known as the {\it doubleton} representations. The degeneracy in the negative branch is finite, but can still be explained exploiting the complex conjugate {\it doubleton} representations. We also determine the spectrum generating algebra for the problem and give the generalization of our results to $d$-dimensions.
		
\vskip 1em
			
\vskip 5pt

\end{quote}
		
\end{titlepage}
%%%%%%%%%%%%%%%%%%%%%%%%%%%%%%%%%%%%%%%%%%%%%%%
%%%%%%%%%%%%%%%%%%%%%%%%%%%%%%%%%%%%%%%%%%%%%%%
%%%%%%%%%%%%%%%%%%%%%%%%%%%%%%%%%%%%%%%%%%%%%%%
\begin{comment}
	\cite{Haaker}, 
	\cite{LiWu}, 
	\cite{Li:2011it} 
	\cite{Li:2012xja} 
	\cite{KaneMele2005} 
	\cite{QiandZhangPhysToday} 
	\cite{Hasan-Kane-10} 
	\cite{Qi-Zhang-11} 
	\cite{SchnyderRFL2008} 
	\cite{Kitaev2008} 
	\cite{RyuSFL2009}
	\cite{QiHZ2008} 
	\cite{EstienneRB2012} 
	\cite{NeupertSRChMRB2012} 
	\cite{Haldane2011} 
	\cite{LiZhangWu2013} 
	\cite{ZhangHu2001} 
	\cite{KarabaliNair2002} 
	\cite{Bernevig2003} 
	\cite{HasebeKimura2003} 
	\cite{Zhang-H-K-1988} 
	\cite{Zhang-1992} 
	\cite{Govil:2013uta} 
	\cite{Govil:2014uwa} 
	\cite{Fernando:2015tiu} 
	\cite{Gunaydin:2016bqx} 
	\cite{Sperling:2018xrm} 
	\cite{Balachandran:2002bj} 
	\cite{Coskun:2016ybb} 	
	
	vonKlitzing:1980pdk, Thouless:1982zz Kohmoto Laughlin Haldane:1988zza
	
	Cappelli:1992yv Iso:1992aa Martinez:1993xv Dirac:1963ta

\end{comment}
%%%%%%%%%%%%%%%%%%%%%%%%%%%%%%%%%%%%%%%%%%%%%%%
%%%%%%%%%%%%%%%%%%%%%%%%%%%%%%%%%%%%%%%%%%%%%%%
%%%%%%%%%%%%%%%%%%%%%%%%%%%%%%%%%%%%%%%%%%%%%%%	
\setcounter{footnote}{0}
\pagestyle{plain} \setcounter{page}{2}
	
\newpage

\section{Introduction}

There is ongoing intense interest in gaining further theoretical insights from new and diverse perspectives on the topological phases of matter discovered in the past decade or so. This is reflected in the vast and diverse literature on the subject \cite{Haaker, KaneMele2005, Bernevig:2006zz, QiandZhangPhysToday, Hasan-Kane-10, Qi-Zhang-11, SchnyderRFL2008, Kitaev2008, RyuSFL2009, QiHZ2008, EstienneRB2012,NeupertSRChMRB2012, Ryu-Takayanagi-2010, Furusaki-etal-2012, LiWu, Li:2011it, Li:2012xja, Asorey:2013wvh}. Integer and fractional quantum Hall (QH) states \cite{vonKlitzing:1980pdk, Tsui:1982yy, Thouless:1982zz, Kohmoto, Laughlin}, which were discovered in early 80's, may be interpreted as early examples, which carry distinct topological numbers distinguishing them from the ordinary states of matter. In the low energy limit, QH system admits an effective field theoretic description in terms of topological Chern-Simons (CS) gauge theory in $2+1$ dimensions \cite{Zhang-H-K-1988,Zhang-1992}. It is also known that QH states belong to a more general class of time reversal (TR) symmetry breaking systems, in which the Hall conductance is quantized (in units of $\frac{e^2}{h}$). More recent interest in the field is driven by the discovery of the new topological phases of matter in two and subsequently in three and higher dimensions, which respect the TR symmetry. First examples of TR-invariant systems in two-dimensions are topological insulators (TIs), which are constructed using the Bloch-wave band structure \cite{KaneMele2005} generalizing the earlier seminal work of Haldane \cite{Haldane:1988zza} by introducing the spin-orbit (SO) interaction and restoring the TR symmetry. In \cite{Bernevig:2006zz} (see also \cite{QiandZhangPhysToday}), Bernevig and Zhang formulated a continuum version of these two-dimensional TR invariant TIs. This is named as the quantum spin Hall effect (QSHE) and can be viewed as consisting of two integer QH states with opposite chirality, for which the charge Hall conductance vanishes, while the spin Hall conductance is quantized (in units of $\frac{e}{2 \pi}$).

New phases of matter with non-trivial topology in three and higher dimensions are also being investigated rigorously from several different directions in the recent past \cite{Hasan-Kane-10, Qi-Zhang-11, SchnyderRFL2008, Kitaev2008, RyuSFL2009, QiHZ2008}. Complementing the investigations of three dimensional TR invariant TIs using the Bloch wave band structure, Li and Wu \cite{LiWu} introduced an interesting continuum model for three and higher dimensional TR invariant TIs. The model proposed by these authors is described by Hamiltonians, in which the charged spin $1/2$ particles are non-minimally coupled to $SU(2) \simeq SO(3)$ and $SO(d)$, Aharanov-Casher type non-abelian gauge fields in three and $d$-dimensions, respectively. These Hamiltonians can also be expressed as that of three and $d$-dimensional harmonic oscillators with a spin-orbit (SO) term at a critical coupling strength, matching the frequency of the harmonic oscillator and has the property that, depending on the sign of the SO term, either the positive or the negative helicity component exhibits flat spectra. The latter is a characteristic of the Landau levels, and therefore the models proposed in \cite{LiWu} may be taken as the generalization of the TR invariant LLs of QSHE \cite{Bernevig:2006zz} to flat higher-dimensional spaces\footnote{These developments may be contrasted to the generalization of QH physics and LLs to higher-dimensional manifolds, where the charged particles are coupled to non-abelian gauge fields that have uniform strength on the given manifold \cite{ZhangHu2001, KarabaliNair2002,Bernevig2003, HasebeKimura2003, Coskun:2016ybb}. These systems are not TR-invariant as they generalize the standard QHE in two dimensions and are used as models of TR breaking TIs \cite{Hasebe:2014nia}.}. In addition to the flat spectra, these model feature other intriguing properties; for instance, in three dimensions LL wave functions satisfy quaternionic analyticity, generalizing the complex analytic property of the ordinary QHE wave functions in the symmetric gauge. However, they do also have features which deviate from the two-dimensional case; lack of full translational symmetry due to the non-abelian form of the gauge field, being one such example. Thus, it is not possible to exploit the Bloch wave function formalism to directly compute the bulk topological index. Nevertheless, authors of \cite{LiWu} have shown that, these models posses gapless helical Dirac surface states, which are robust against the TR invariant perturbations and therefore allow for the interpretation of the models as TR preserving TIs with a ${\mathbb Z}_2$ topological invariant. To be somewhat more concrete, as exhibited in \cite{LiWu, LiWuSup} through a numerical calculation in the three-dimensional case under open boundary conditions, flat spectra acquires dispersion and this clearly reveals the presence of the surfaces states. The spectrum of the latter can be linearized around the Fermi angular momentum and the surface Hamiltonian can be given in terms of the Dirac operator on the spherical boundary. This analysis easily generalizes to $d$-dimensions with surface states governed by the Dirac Hamiltonian on $S^{d-1}$. Each fully occupied LL contributes one branch of helical surface modes at the spherical boundary and the model is topologically ${\mathbb Z}_2$ non-trivial if an odd number of LLs are filled. A related model breaking the full rotational symmetry, but with similar findings is discussed in \cite{ Li:2012xja}, while the square root problem of LL for Dirac fermions is presented in \cite{Li:2011it}. 

In \cite{Haaker} properties and structure of the wave functions of the three dimensional TI system are explored from an algebraic perspective and an underlying non-compact extended dynamical symmetry group that completely accounts for both the infinite degeneracy of the LLs in the positive helicity and the finite degeneracy in the negative helicity component is determined. In practice, authors of \cite{Haaker} reveal two vector operators involving the total angular momentum (orbital and spin) and commuting with the Hamiltonian, which resemble the Runge-Lenz vector of the Kepler problem \cite{Kepler}. Appropriately scaled form of these vector operators on the eigenstates of the Hamiltonian and the generators of the total angular momentum, span the noncompact Lie algebra $so(3,2)$. The Dirac {\it Singleton} \cite{Dirac:1963ta, FF} is a well-known unitary irreducible representation (UIR) of $so(3,2)$ and plays the central role in explaining the aforementioned infinite and finite degeneracies in the spectrum of this model. It may be noted that, knowledge on the algebraic structure of the TI wave functions could be very useful, since it could allow for a deeper understanding of {\it i.} the underlying geometric features and {\it ii.} the algebraic organization of the surface states. To elaborate on the first item, it is  speculated in \cite{Haaker} that the underlying geometry of LLs in three dimensions could be related to some radially deformed form of $AdS_4$, rather than the flat space, since the extended symmetry group can be thought as a ``deformed" $SO(3,2)$, recalling that $SO(3,2)$ is the exact isometry group of $AdS_4$, while for the second we may recall the $W_{\infty}$ symmetry encountered in certain QH phases, which accounts for the incompressibility of the QH droplet and algebra of the edge states \cite{Ezawa, Cappelli:1992yv, Iso:1992aa, Martinez:1993xv}, whose generalization to TR invariant TIs would be very interesting. 

In the present work, we extend the results of \cite{Haaker} to four and subsequently to $d$-dimensional models. Introducing the appropriate vector operators commuting with the Hamiltonian and obtaining their suitably scaled form acting on the energy eigenkets, the extended dynamical symmetry group in four dimensions is identified as $SO(4,2)$. We show in full detail how the infinite degeneracy of the energy spectrum in the positive helicity branch can be explained in terms of the discrete UIRs of $SO(4,2)$, which are also known as the {\it doubletons} \cite{Govil:2013uta, Govil:2014uwa,  Sperling:2018xrm}. The finite degeneracy in the negative branch is also understood using the complex conjugate {\it doubleton} representations. In section $4$, we give the generalization of our analysis to the models in $d$-dimensions and reveal that the extended symmetry group is $SO(d,2)$, while the degeneracies are explained using the discrete series UIRs of $SO(d,2)$, which essentially generalize the {\it singleton} and the {\it doubleton} representations in $d=3$ and $d=4$ to higher odd and even dimensions, respectively \cite{Fernando:2015tiu, Gunaydin:2016bqx}.

Our results allow us to gain a broader perspective into the structure of the $d \geq 3$-dimensional LLs. In particular, we observe that the operator $A = \sum_{a<b} L_{ab} \Gamma_{ab} + \frac{d-1}{2}$, i.e. the spin-orbit coupling term (up to a constant shift) identifies with the $U(1)$ generator of $SO(d,2)$ w.r.t which the Lie algebra $so(d,2)$ has a three-graded decomposition. We may recall that the $SO(d,2)$ is the conformal group for $d$-dimensional Minkowski space-time and this particular $U(1)$ generator is identified as the conformal Hamiltonian and its spectrum as the conformal energy. The latter may also be interpreted as the $AdS_{d+1}$ energy via AdS/CFT duality \cite{Fernando:2015tiu, Gunaydin:2016bqx}. Thus, we have the picture that the extended dynamical symmetry $SO(d,2)$ of $d$-dimensional TR invariant TIs reveals the algebraic organization of the LL states w.r.t to the underlying ``deformed" $AdS_{d+1}$ geometry, where the eigenvalues of $A$ shift by $\pm 1$ under the action of $so(d,2)$ ladder operators, that are organized in accord with its three-grading. As we have noted in the preceding paragraph, the boundary Hamiltonian is given in terms of the Dirac operator on $S^{d-1}$ (for a discussion of Dirac operators on $S^{d-1}$ \cite{Balachandran:2002bj} can be consulted). It can therefore be expressed in terms of the operator $A$ instead. Although $SO(d,2)$ can no longer be considered as the extended dynamical symmetry group once the open boundary conditions are imposed, to the extent that it may be applied, it may serve as an effective spectrum generating algebra for the surface states as the eigenvalues of the boundary Hamiltonian shift by $\pm \frac{v_F}{R_0}$ under the action of $so(d,2)$ ladder operators, $v_F$ being the Fermi velocity and $R_0$ the radius at the open boundary. We present a brief discussion of these observations in the final section of the paper.      
 
\section{The Model for Four-Dimensional LLs}

\subsection{Basics and Preliminary Remarks}

We may launch our discussion starting with the Hamiltonian of a four-dimensional ($4D$) harmonic oscillator coupled to an Aharanov-Casher type $SO(4)$ gauge field $G_a = 2 m \omega r_b S_{ab}$ in the form \cite{LiWu}
\be
H = \frac{1}{2m} (p_a - G_a)^2 - m \omega^2 r_a^2 \,,
\label{H1}
\ee
which may be expressed as the Hamiltonian for a simple harmonic oscillator (SHO) with the spin-orbit (SO) term at the coupling strength $\omega$ matching the SHO frequency as 
\be
H = \frac{p_a^2}{2m} + \frac{1}{2} m \omega^2 r_a^2 - \omega \sum_{a <b = 1 }^{4}  L_{ab} \Gamma_{ab}  \,.
\label{H2}
\ee
In this expression $L_{ab} : = r_a p_b - r_b p_a \,, (a,b= 1,\cdots,4)$ are the orbital angular momentum operators, while $\Gamma_{ab}$ are proportional to the spin operator $S_{ab}$ in $4$-dimensions, as will be explicitly defined in what follows. Setting $\hbar = 1$, we may write the momentum operator as $p_a = -i \partial_a$. In terms of the representation theory of $SO(4)$, $L_{ab}$ carries the $(l, 0)$ irreducible representation (IRR) of $SO(4)$, while $S_{ab}$ carries the direct sum representation $(1/2, 1/2) \oplus (1/2, -1/2)$ (IRRs are given in the highest weight i.e. the Gelfand-Zeitlin notation). To be more concrete, let us introduce the $4$-dimensional Euclidean $\gamma$-matrices, $\gamma_a \,, (a= 1,\cdots,4)$ with the anti-commutation relations $\lbrace \gamma_a \,, \gamma_b \rbrace = 2 \delta_{ab}$. We may choose them to be of the form 
\beqa
&&\gamma_i = 
\left (
\begin{array}{cc}
	0 & - i \sigma_i  \\ 
	i \sigma_i	& 0
\end{array} 
	\right )
  \,, \quad 
\gamma_4 = \left (
\begin{array}{cc}
0 & 1 \\ 
1 & 0
\end{array} 
 \right )
 \,,\quad (i=1,2,3) \\
&&\gamma_5 = \gamma_1 \gamma_2 \gamma_3 \gamma_4 =
\left (
\begin{array}{cc}
1 & 0 \\ 
0 & -1
\end{array}
\right ) \,.
\eeqa
Spin operator, $S_{ab}$, may be expressed as
\be
S_{ab} := \frac{1}{2} \Gamma_{ab} : = -\frac{i}{4} \lbrack \gamma_a \,, \gamma_b \rbrack \,, \quad 
S_{ab} = \left (
\begin{array}{cc}
S_{ab}^+ & 0  \\ 
0	& S_{ab}^- 
\end{array} 
\right ) \,, \quad S_{ab}^\pm = (S_{ij}, \mp \frac{1}{2} \sigma_i) =  ( \frac{1}{2}\varepsilon_{ijk} \sigma_k, \mp \frac{1}{2} \sigma_i) \,.
\ee

Total angular momentum is given as $J_{ab} = L_{ab} + S_{ab}$ and has the IRR content given by the decomposition of the product $(l,0)\otimes [(1/2,1/2) \oplus (1/2,-1/2)]$ as
\be
\left (l+\frac{1}{2},\frac{1}{2} \right ) \oplus \left(l-\frac{1}{2}, \frac{1}{2}  \right ) \oplus \left (l+\frac{1}{2}, -\frac{1}{2}  \right ) \oplus \left (l-\frac{1}{2}, - \frac{1}{2}  \right ) \,.
\label{irr1}
\ee

$SO(4)$ commutation relations are given in terms of generic generators $M_{ab}$ as 
\be
\lbrack M_{ab}, M_{cd} \rbrack = i (\delta_{ac} M_{bd} + \delta_{bd} M_{ac} - \delta_{ad} M_{bc} - \delta_{bc} M_{ad}) \,.
\label{genericcom1}
\ee
$L_{ab}$, $S_{ab}$ and $J_{ab}$ satisfy (\ref{genericcom1}). 

The Hamiltonian commutes with the total angular momentum operator $J_{ab}$. Its spectrum and eigenfunctions are given in \cite{LiWu, LiWuSup}. We briefly present some details in order to be self-contained and prepare for the developments that follow. Spectrum of the pure $4D$ SHO is given as $E_{4D \,, SHO} = \omega (2 n + \ell +2)$ and the corresponding energy eigenfunctions are of the form $\Psi(r, \theta, \phi, \psi) = R_{n \ell}(r) \, Y^l_{m_L m_R}(\theta, \phi, \psi)$, where $R_{n \ell}(r) = r^l e^{- \frac{1}{2} m \omega r^2} F(-n, l+2, m \omega r^2)$ with $n \in {\mathbb Z}_+$ is the radial wave function and $Y^\ell_{m_L m_R}(\theta, \phi, \psi)$ are the spherical harmonics in four dimensions. 

Eigenvalues of the SO term can easily be worked out using the eigenvalues of the Casimir operators for the IRRs appearing in (\ref{irr1}).
%\footnote{Eigenvalues of the Casimir operators for the relevant representations of $SO(4)$ and $SO(d)$ are given in the appendix}. 
We have    
\beqa
\sum_{a<b} L_{ab} S_{ab}^\pm  &=& \frac{1}{2} (J_{ab}^2 - L_{ab}^2- S_{ab}^{\pm 2}) 
=  
\begin{dcases}
\quad \frac{l}{2} & \mbox{on} \left( l +\frac{1}{2} \,, \pm \frac{1}{2} \right) \,, \quad \mbox{i.e. spin} \, \uparrow \, \\
-\frac{l+2}{2} & \mbox{on} \left( l - \frac{1}{2} \,, \pm \frac{1}{2} \right) \,, \quad \mbox{i.e. spin} \, \downarrow	
\end{dcases}	
.
\label{SOterm_spec}
\eeqa

Spectrum of the Hamiltonian in (\ref{H2}) then follows as
\be
E =
\begin{dcases}
2 \omega (n+1) \,, & \mbox{spin $\uparrow$}	\\
2 \omega (n + l + 2) \,, & \mbox{spin $\downarrow$}	
\end{dcases} \,,
\label{spec1}	
\ee	
from which we observe that the spin up (positive SO branch) part has flat spectrum, i.e. it is independent of the orbital angular momentum $l$, and leads to an infinite degeneracy at each energy level. Spin down (negative SO branch) part of the spectrum is also degenerate, but not infinitely so. In the ensuing sections our main focus will be explaining the reason underlying this degeneracy. It is useful to note that the infinite degeneracy of the positive SO branch is a direct consequence of the critical SO coupling strength which matches with the SHO frequency $\omega$; in particular, changing the sign of the SO term in the Hamiltonian would flip the spectrum of the positive and negative SO branches, making the latter infinitely degenerate instead. Corresponding wave functions are $R_{nl}(r) {\cal Y}_{l; m_L \, m_R}^{l \pm \frac{1}{2}}(\theta, \phi, \psi)$ where $R_{n l}(r)$ is the same as before, while ${\cal Y}_{l; m_L \, m_R}^{l \pm \frac{1}{2}}(\theta, \phi, \psi)$ are the spin spherical harmonics in four dimensions. 

In analogy with the discussion of \cite{Haaker} in three dimensions, we find it useful to introduce the operator 
\be
A = \sum_{a<b} L_{ab} \Gamma_{ab} + \frac{3}{2} \,.
\label{opA}
\ee
Using equation (\ref{SOterm_spec}) eigenvalues of $A$ can be simply written as $l^\prime := l + \frac{3}{2}$ for spin up and $l^\prime := - l - \frac{1}{2}$, for spin down, respectively. Thus, we have $l^\prime = \pm \frac{3}{2} \,, \pm \frac{5}{2} \,, \cdots$. Evidently, $A$ commutes with the Hamiltonian and  therefore its eigenvalues $l^\prime$ can be used in labeling the energy eigenstates. Since $SO(4) \simeq SU(2) \times SU(2)$, we can introduce $SU(2)$-left and $SU(2)$-right generators for the total angular momentum $J_{ab}$ as 
\be
L_i = \frac{1}{2} ( \frac{1}{2} \varepsilon_{ijk} J_{jk} + J_{i4}) \,, \quad R_i = \frac{1}{2} ( \frac{1}{2} \varepsilon_{ijk} J_{jk} - J_{i4}) \,,
\ee
with the commutation relations 
\be
\lbrack L_i \,, L_j \rbrack = i \varepsilon_{ijk} L_k \,, \quad \lbrack R_i \,, R_j \rbrack = i \varepsilon_{ijk} R_k \,, \quad
\lbrack L_a ,, R_b \rbrack = 0 \,. 
\ee
Conventionally, generators of the Cartan subgroup of $SO(4)$ are taken as $(J_{12}, J_{34})$, while for $SU(2) \times SU(2)$ they are taken as
\be
(L_3 , R_3) = \left ( \frac{1}{2}(J_{12} + J_{34}) \,, \frac{1}{2} (J_{12} - J_{34}) \right ) \,.
\label{su2su2}
\ee
From (\ref{irr1}) and (\ref{su2su2}) we see that the fundamental representations $(1/2, 1/2)$ and $(1/2, -1/2)$ correspond respectively to $(L_i, R_i) \equiv  (0\,, \frac{\sigma_i}{2}) $ and  $(L_i, R_i) \equiv  (\frac{\sigma_i}{2} \,, 0)$. In the $SU(2) \times SU(2)$ irreducible representation notation, $(j_1, j_2)$, these are labeled as $(0,1/2)$ and $(1/2,0)$, respectively. 

We choose to label the eigenstates of the Hamiltonian in terms of the principal quantum number $n$ and the eigenvalues $l^\prime, m_L, m_R$ of $A$, $L_3$ and $R_3$,  and denote, in the Dirac notation, these states as $|n, l^\prime, m_L, m_R \rangle$ with
\beqa
A |n, l^\prime, m_L, m_R \rangle &=& l^\prime |n, l^\prime, m_L, m_R \rangle \,, \nn  \\
L_3 |n, l^\prime, m_L, m_R \rangle &=& m_L |n, l^\prime, m_L, m_R  \rangle \,,  \\
R_3 |n, l^\prime, m_L, m_R \rangle &=& m_R |n, l^\prime, m_L, m_R  \rangle \,. \nn
\eeqa

In terms of the $SU(2) \otimes SU(2)$ representation labels the direct sum representation in (\ref{irr1}) reads 
\be
\left ( \frac{l +1}{2} \,, \frac{l}{2} \right) \oplus \left ( \frac{l}{2} \,, \frac{l-1}{2} \right) \oplus
\left ( \frac{l}{2} \,, \frac{l+1}{2} \right) \oplus \left ( \frac{l-1}{2} \,, \frac{l}{2} \right) \,.
\ee
From this, we immediately infer that $|m_L| \leq \frac{l+1}{2}$ and $|m_R| \leq \frac{l}{2}$, and $|m_L| \leq \frac{l}{2}$ and $|m_R| \leq \frac{l-1}{2}$ respectively, for the right chiral representations (i.e. first two summands in (\ref{irr1})). In terms of the eigenvalues of $l^\prime$ of $A$, we have the range of eigenvalues for $m_L$ and $m_R$ expressed as
\be
|m_L|\leq \left ( \frac{|l^\prime|}{2} - \frac{1}{4} \right) \,, \quad  |m_R|\leq \left ( \frac{|l^\prime|}{2} - \frac{3}{4} \right) \,.
\label{intmlmr} 
\ee
As for the range of values for $m_L$ and $m_R$ in the left chiral representations, we simply interchange $m_L$ and $m_R$ in (\ref{intmlmr}).
 
Trading the label $l$ for $l^\prime$, we can express the spectrum in (\ref{spec1}) as 
\be
E =
\begin{dcases}
	2 \omega (n+1) \,, & \mbox{spin $\uparrow$}	\\
	2 \omega (n - l^\prime + \frac{3}{2}) \,, & \mbox{spin $\downarrow$}	
\end{dcases} \,.	
\label{spec2}
\ee	

\subsection{Extended Dynamical Symmetries}

In order to understand the infinite and the finite degeneracies of the positive and negative helicity branches of the spectrum (\ref{spec2}), we will reveal this $4D$ model has an extended non-compact dynamical symmetry group. To do so, working from now on with $m=1$ and $\omega= \frac{1}{2}$, generalizing the approach of \cite{Haaker}, we introduce two Hermitian vector operators commuting with the Hamiltonian $H$ and involve, in addition to the coordinates and momenta, the total angular momentum and the SO operator $A$. Explicitly, they are in the form
\beqa
M_a &=& \frac{1}{4} (r_a A + A r_a) + \frac{1}{2} (p_b J_{ab} + J_{ab} p_b) \,, \nn \\
N_a &=& \frac{1}{2} (p_a A + A p_a) - \frac{1}{4} (r_b J_{ab} + J_{ab} r_b) \,.
\label{vecopsMN}
\eeqa 
It can be straightforwardly demonstrated that 
\be 
\lbrack M_a \,, H \rbrack = 0 \,, \quad \lbrack N_a \,, H \rbrack = 0 \,,
\ee
and $M_a$ and $N_a$ transform as vectors under the adjoint action of $J_{ab}$:
\beqa
ad J_{ab} \, M_c &:=& \lbrack J_{ab} \,, M_c \rbrack = i \delta_{ac} M_b - i \delta_{bc} M_a \,, \nn \\
ad J_{ab} \, N_c &: =& \lbrack J_{ab} \,, N_c \rbrack = i \delta_{ac} N_b - i \delta_{bc} N_a \,,
\label{adjvector}
\eeqa
by direct calculation. It is also useful to note that the commutators of these vector operators with $A$, take the form
\be
\lbrack A \,, M_a \rbrack = - i N_a \,, \quad \lbrack A \,, N_a \rbrack = i M_a \,.
\label{AMN}
\ee 

A set of long and rather tedious calculations yield the commutation relations for the operators $M_a$ and $N_a$ as
\beqa
\lbrack M_a \,, M_b \rbrack  &=& - 2 i J_{cd} \left ( \delta_{ac} \delta_{bd} \left (H + \frac{3}{2} A - 1 \right) + \frac{1}{8} \varepsilon_{abcd} \gamma_5 \right) \,, \nn \\
\lbrack N_a \,, N_b \rbrack  &=& - 2 i J_{cd} \left ( \delta_{ac} \delta_{bd} \left (H + \frac{3}{2} A - 1 \right) + \frac{1}{8} \varepsilon_{abcd} \gamma_5 \right) \,, \nn \\
\lbrack M_a \,, N_b \rbrack  &=&  2 i \delta_{ab} A \left  (H + \frac{3}{2} A - 1 \right) + i J_{ac}J_{bc} - i \delta_{ab} \sum_{c<d} J_{cd}^2 \,, \nn \\
&=& 2 i \delta_{ab} \left (  A (H + \frac{3}{2} A - 1 ) - \frac{1}{2} A^2 + \frac{3}{8} \right) + i J_{ac}J_{bc} \,. 
\label{comMN}
\eeqa
We have used, $\sum_{c<d} J_{cd}^2 =  A^2 - \frac{3}{4}$ to express the second line of the last commutator in (\ref{comMN}). We may form the following linear combinations of $M_a$ and $N_a$
\beqa
K^1_\pm &:=& \frac{1}{\sqrt2}(M_1 \pm i M_2 \mp i N_1 + N_2) \,, \nn \\ 
K^2_\pm &:=& \frac{1}{\sqrt2} (M_1 \mp i M_2 \mp i N_1 - N_2) \,,  \\
K^3_\pm &:=& \frac{1}{\sqrt2}( M_3 \pm i M_4 \mp i N_3 + N_4 )\,, \nn \\ 
K^4_\pm &:=& \frac{1}{\sqrt2}(M_3 \mp i M_4 \mp i N_3 - N_4 ) \,, \nn
\label{KMN}
\eeqa 
which fulfill the commutation relations 
\be
\begin{array}{lll}	
\lbrack A \,,K^1_\pm \rbrack = \pm   K^1_\pm \,, \quad &	\lbrack L_3 \,,K^1_\pm \rbrack = \pm  \frac{1}{2} K^1_\pm \,, \quad & \lbrack R_3 \,, K^1_\pm \rbrack = \pm \frac{1}{2} K^1_\pm \,, \\
\lbrack A \,,K^2_\pm \rbrack = \pm  K^2_\pm \,, \quad &	\lbrack L_3 \,, K^2_\pm \rbrack = \mp    \frac{1}{2} K^2_\pm  \,, \quad & \lbrack R_3 \,,K^2_\pm \rbrack = \mp \frac{1}{2} K^2_\pm \,, \\
\lbrack A \,,K^3_\pm \rbrack = \pm   K^3_\pm \,, \quad &	\lbrack  L_3 \,, K^3_\pm \rbrack = \pm \frac{1}{2} K^3_\pm \,, \quad & \lbrack R_3 \,,  K^3_\pm \rbrack = \mp \frac{1}{2} K^3_\pm \,, \\
\lbrack A \,,K^4_\pm \rbrack = \pm  K^4_\pm \,, \quad &	\lbrack  L_3 \,, K^4_\pm \rbrack = \mp \frac{1}{2} K^4_\pm \,, \quad & \lbrack R_3 \,, K^4_\pm \rbrack = \pm \frac{1}{2} K^4_\pm \,.	
\label{extendedcom1}
\end{array}
\ee   
Comparison of these commutation relations with those of $so(4,2)$ roots and Cartan generators as given in the next section in (\ref{comso42part1}) and (\ref{comso42part2}) suggests a correspondence between $L_\pm$, $R_\pm$, $K_\pm^i$, ($i=1,2,3,4$) and the roots $E_{\pm (e^i \pm e^j)}$ of $so(4,2)$. This is obvious for the $so(4)= su(2)_L \oplus su(2)_R$ subalgebra. For this proposed correspondence the operator $A$ needs to be identified with a particular Cartan generator of $so(4,2)$, as we will lay out in detail in the following section. Nevertheless, the commutation relations among $K_\pm^i$, as inferred from those of $M_a$ and $N_a$ in (\ref{comMN}) include nonlinear terms in $H$ and $A$ and does not immediately fit into the $so(4,2)$ commutation relations. For instance, we find
\beqa
\lbrack K_-^1 \,, K_+^1 \rbrack &=& 4 (J_{12} + A) \left  (H + \frac{3}{2} A - 1 \right) +  J_{34} -  \sum_{c<d} J_{cd}^2 +  (J_{12}+J_{34})(J_{12}-J_{34}) \,, \nn \\
&=& 4 (L_3 + R_3 + A) \left  (H + \frac{3}{2} A - 1 \right) + (L_3 - R_3) -  A^2 + \frac{3}{4} + 4 L_3 R_3 \,.
\label{KKcom1}
\eeqa
Such complications are encountered in several different contexts, for instance in the Kepler problem in identifying $SO(4)$ as the extended dynamical symmetry group of the Hydrogen atom \cite{Kepler}. It was also faced in the $3D$ case treated in \cite{Haaker}. This issue can be remedied by appropriately scaling the operators $K_\pm^i$ acting on the energy eigenstates $|n, l^\prime, m_L, m_R \rangle$. It turns out that the suitable scalings of $K_\pm^i$ can be obtained by exploiting the operator
\be
S = 4 ( H + A - \frac{3}{2}) \,,
\ee
as we will discuss in detail in the next section.

Then, $A, L_i, R_i$, and 
\be
\frac{1}{\sqrt{S}}K_+^i \,, \quad K_-^i \frac{1}{\sqrt{S}} \,,
\ee
generate the non-compact group $SO(4,2)$ on the energy eigenstates $|n, l^\prime, m_L, m_R \rangle$. Thus, we identify $SO(4,2)$ as the extended dynamical symmetry of the model described by the Hamiltonian in (\ref{H2}). In the next section, by providing the details of this result, we show how the infinite-fold degeneracy of the positive helicity branch can be labeled in terms of a particular discrete UIR of $SO(4,2)$. We will also see how the finite degeneracy of the negative helicity branch is explained using the same machinery and a related UIR.

\section{Discrete UIRs of $SO(4,2)$ and the Degenerate LL Spectrum}

Hermitian generators $M_{\mu \nu}$ of the $so(4,2)$ Lie algebra satisfy the commutation relations 
\be
\lbrack M_{\mu \nu} \,, M_{\rho \sigma} \rbrack = i (\eta_{\mu \rho} M_{\nu \sigma} + \eta_{\nu \sigma} M_{\mu \rho} - \eta_{\mu \sigma} M_{\nu \rho} - \eta_{\nu \rho} M_{\mu \sigma}) \,. 
\ee
Here, we use the metric convention $\eta_{\mu \nu} = \mbox{diag}(1,1,1,1,-1,-1)$, $(\mu \,, \nu= 1 \,, \cdots \,, 6)$.

Cartan subalgebra of $so(4,2)$ is generated by $(H_1,H_2, H_3) \equiv (M_{12}, M_{34}, M_{56})$. Introducing the three-component unit vectors $e^1,e^2,e^3$ with $(e^i)_j = \delta^i_j$, the roots may be expressed as
\be
E_{\pm(e^1-e^2)} \,, \quad E_{\pm(e^2-e^3)} \,, \quad E_{\pm(e^2+e^3)} \,, \quad E_{\pm(e^1+e^2)} \,, \quad E_{\pm(e^1-e^3)} \,, \quad E_{\pm(e^1+e^3)} \,.
\label{rootsso42}      
\ee
It is useful to introduce the notation $E_{\pm \alpha^\mu}$ for the roots, with the labels given as $\pm \alpha^\mu := \pm(e^i \pm e^j)$ with $i<j$.  In a standard short-hand notation of the Cartan-Weyl basis, commutation relations among the generators can be compactly expressed as \cite{Fuchs}
\beqa
&&\lbrack  H_i \,, H_j \rbrack = 0 \,, \quad \lbrack H_i \,, E_{\alpha^\mu} \rbrack = \alpha_{i}^\mu E_{\alpha^{\mu}} \,, \nn \\
&&\lbrack E_{\alpha^\mu} \,, E_{\alpha^\nu} \rbrack = 
\begin{dcases}
N_{\alpha^\mu \alpha^\nu} E_{\alpha^\mu + \alpha^\nu} \,, \quad \mbox{if $\alpha^\mu + \alpha^\nu$ is a root} \\
(E_{\alpha^\mu} \,,E_{-\alpha^\mu} ) \alpha^\mu_{i} H_i \,, \, \mbox{if $\alpha^\mu + \alpha^\nu = 0$, and sum over $i$ is implied}, \\
0 \,, \quad \mbox{otherwise}
\end{dcases}
\eeqa
where $( E_{\alpha^\mu} \,, E_{\alpha^\nu} )  = \frac{1}{2} \mbox{Tr} \, E_{\alpha^\mu} E_{\alpha^\nu}$ and the normalized traces are given by 
\be
\mbox {Tr} \, M_{\mu \nu}M_{\rho \sigma} = 2 \eta_{\mu \rho} \eta_{\nu \sigma} - 2 \eta_{\mu \sigma} \eta_{\nu \rho} \,. 
\ee 
Cartan subalgebra and the roots form the Cartan-Weyl basis for the fifteen generators of $so(4,2)$. Roots in (\ref{rootsso42}) can be expressed as linear combinations of the $M_{\mu \nu}$ as follows:
\beqa
E_{\pm(e^1-e^2)} &=& \frac{1}{2} (\mp i M_{13}+M_{23}-M_{14} \mp i M_{24}) \,, \nn \\
E_{\pm(e^2-e^3)} &=& \frac{1}{2} (M_{35} \pm i M_{45}-M_{46} \pm i  M_{36}) \,, \nn \\
E_{\pm(e^2+e^3)} &=& \frac{1}{2} (M_{35} \pm i M_{45}+ M_{46} \mp i M_{36}) \,, \nn \\
E_{\pm(e^1+e^2)} &=& \frac{1}{2} (\pm i M_{23} + M_{13} \pm i M_{14} - M_{24}) \,, \\
E_{\pm(e^1-e^3)} &=& \frac{1}{2} (\pm i M_{25} - M_{26} + M_{15} \pm i M_{16}) \,, \nn \\
E_{\pm(e^1+e^3)} &=& \frac{1}{2} (\mp i M_{25} - M_{26} - M_{15} \pm i M_{16}) \,. \nn
\eeqa

The subalgebra $so(4)$ of $so(4,2)$ has the Cartan generators $H_1, H_2$. In the $so(4) = su(2) \times su(2)$ basis, Cartan generators can be taken as $L_3$ and $R_3$, which are given in terms of $H_1$ and $H_2$ as
\be
L_3 = \frac{1}{2} (H_1 + H_2) \,, \quad R_3 = \frac{1}{2} (H_1 - H_2) \,.
\ee  
The relevant part of the commutation relations among the generator can be summarized as
\be
\begin{array}{ll}	
	\lbrack L_3 \,, E_{\pm(e^1-e^2)} \rbrack = 0 \,, \quad & \lbrack R_3 \,, E_{\pm(e^1-e^2)} \rbrack = \pm E_{\pm(e^1-e^2)} \,, \\
	\lbrack  L_3 \,, E_{\mp(e^2-e^3)} \rbrack = \mp \frac{1}{2} E_{\mp(e^2-e^3)} \,, \quad & \lbrack R_3 \,, E_{\mp(e^2-e^3)} \rbrack = \pm \frac{1}{2} E_{\mp(e^2-e^3)} \,, \\
	\lbrack  L_3 \,, E_{\pm(e^2+e^3)} \rbrack = \pm \frac{1}{2} E_{\pm(e^2+e^3)}  \,, \quad & \lbrack R_3 \,, E_{\pm(e^2+e^3)} \rbrack = \mp \frac{1}{2} E_{\pm(e^2+e^3)}  \,, \\
	\lbrack L_3 \,, E_{\pm(e^1+e^2)} \rbrack = \pm  E_{\pm(e^1+e^2)}  \,, \quad & \lbrack R_3 \,, E_{\pm(e^1+e^2)} \rbrack = 0 \,, \\
	\lbrack  L_3 \,, E_{\mp(e^1-e^3)} \rbrack = \mp \frac{1}{2} E_{\mp(e^1-e^3)} \,, \quad & \lbrack R_3 \,,  E_{\mp(e^1-e^3)} \rbrack = \mp \frac{1}{2} E_{\mp(e^1-e^3)} \,, \\
	\lbrack  L_3 \,, E_{\pm(e^1+e^3)} \rbrack = \pm \frac{1}{2} E_{\pm(e^1+e^3)} \,, \quad & \lbrack R_3 \,,  E_{\pm(e^1+e^3)} \rbrack = \pm \frac{1}{2} E_{\pm(e^1+e^3)} \,,	
\end{array}
\label{comso42part1}
\ee

\be
\begin{array}{ll}
\lbrack H_3 \,, E_{\pm(e^1-e^2)} \rbrack = 0 \,, \quad & \lbrack E_{(e^1-e^2)} \,,  E_{-(e^1-e^2)} \rbrack = H_1 -H_2 \,, \\
\lbrack H_3 \,, E_{\mp(e^2-e^3)} \rbrack = \pm E_{\mp(e^2-e^3)} \,, \quad& \lbrack E_{(e^2-e^3)} \,,  E_{-(e^2-e^3)} \rbrack = - H_2 + H_3 \,, \\
\lbrack H_3 \,, E_{\pm(e^2+e^3)} \rbrack = \pm E_{\pm(e^2+e^3)}  \,, \quad& \lbrack E_{(e^2+e^3)} \,,  E_{-(e^2+e^3)} \rbrack = - H_2 - H_3 \,, \\
\lbrack H_3 \,, E_{\pm(e^1+e^2)} \rbrack = 0 \,,\quad & \lbrack E_{(e^1-e^3)} \,,  E_{-(e^1-e^3)} \rbrack = - H_1 + H_3 \,, \\
\lbrack H_3 \,, E_{\mp(e^1-e^3)} \rbrack = \pm E_{\mp(e^1-e^3)} \,,\quad & \lbrack E_{(e^1+e^3)} \,,  E_{-(e^1+e^3)} \rbrack = - H_1 - H_3 \,, \\
\lbrack H_3 \,, E_{\pm(e^1+e^3)} \rbrack = \pm E_{\pm(e^1+e^3)} \,,\quad & \lbrack E_{(e^1+e^2)} \,,  E_{-(e^1+e^2)} \rbrack = H_1 + H_2 \,. 
\label{comso42part2}
\end{array}
\ee

We are interested in the discrete UIRs of the $so(4,2)$ Lie algebra and the corresponding Lie group $SO(4,2)$. These are usually called the {\it doubletons} \cite{Govil:2013uta,Govil:2014uwa} in the literature and they are bounded from below. They can be built via their lowest weight states.  In order to construct these UIRs, we take advantage of the maximally compact subgroup $SU(2)_L \otimes SU(2)_R \otimes U(1)$ of $SO(4,2)$, which has the same Cartan subalgebra as that of $SO(4,2)$. The $U(1)$ part here is generated by $H_3 = M_{56}$, and it is usually called the conformal Hamiltonian in the literature \cite{Govil:2013uta,Govil:2014uwa}, while the $SU(2)_L \otimes SU(2)_R$ is generated by
\be
L_{i} = \frac{1}{2} \left ( \frac{1}{2} \varepsilon_{ijk} M_{jk} + M_{i4} \right) \,, \quad R_{i} = \frac{1}{2} \left ( \frac{1}{2} \varepsilon_{ijk} M_{jk} - M_{i4} \right) \,, \quad i = 1,2,3 \,.
\ee
With respect to $H_3$, the Lie algebra $so(4,2)$ admits the three-graded decomposition \cite{Govil:2013uta,Govil:2014uwa}
\be
so(4,2) \equiv {\cal L}^+ \oplus {\cal L}^0 \oplus {\cal L}^- \,,
\label{threegraded1}
\ee
where ${\cal L}^0$ stands for the maximally compact subalgebra $su(2)_L \oplus su(2)_R \oplus u(1)$ and ${\cal L}^\pm$ contain the remaining generators, with the three-grading defined as
\be
\lbrack {\cal L}^0 \,, {\cal L}^\pm \rbrack =  {\cal L}^\pm \,, \quad \lbrack H_3 \,, {\cal L}^\pm \rbrack = \pm {\cal L}^\pm \,.
\label{threegrading1}
\ee  
Out of the six pairs of roots $E_{\pm (e^i \pm e^j)}$, we have $L_\pm = E_{\pm (e^1 + e^2)}$ which generate $su(2)_L$ together with $L_3$ and $E_{\pm (e^1 - e^2)}$ $su(2)_R$ together with $R_3$, while the remaining four pair of roots in ${\cal L}^\pm$ transform as a vector, i.e. in the IRR $(\frac{1}{2}, \frac{1}{2})$ of $su(2)_L \oplus su(2)_R$ as it is already implied by the three-graded decomposition given in (\ref{threegraded1}). We note that these are complex vectors since $E_{\pm (e^i \pm e^j)}^\dagger = E_{\mp (e^i \pm e^j)}$.

In order to proceed, we may introduce the four pairs of annihilation and creation operators, which are split into two ``colors", namely $a$'s and $b$'s as
\be
\lbrack a_\alpha \,, a_\beta^\dagger \rbrack = \delta_{\alpha \beta} \,, \quad \lbrack b_\alpha \,, b_\beta^\dagger \rbrack = \delta_{\alpha \beta} \,, \quad \alpha,\beta = 1,2 \,.
\ee
In terms of these operators, $SU(2)_L$ and $SU(2)_R$ generators can be built in the form
\be
L_{\alpha \beta} = a_\alpha^\dagger a_\beta - \frac{1}{2} {\delta}_{\alpha \beta} {\hat N}_a \,, \quad R_{\alpha \beta}= b_\alpha^\dagger b_\beta - \frac{1}{2} {\delta}_{\alpha \beta} {\hat N}_b \,, 
\ee
where ${\hat N}_a = a_i^\dagger a_i$ and ${\hat N}_b = b_i^\dagger b_i$ are the number operators in the colors $a$ and $b$. ${\cal L}^+$ and ${\cal L}^-$ are spanned by $a_i^\dagger b_j^\dagger$ and $a_i b_j$, respectively.  There is indeed a one to one correspondence between the roots $E_{\pm (e^i \pm e^j)}$ and $a_i^\dagger b_j^\dagger$ and $a_i b_j$, which can be given explicitly as
\be
\begin{array}{ll}
a_1^\dagger b_1^\dagger  \equiv E_{e^1+e^3} \,, \quad & a_1 b_1  \equiv  E_{-(e^1+e^3)} \,, \\
a_1^\dagger b_2^\dagger  \equiv  E_{e^2+e^3} \,, \quad & a_1 b_2 \equiv  E_{-(e^2+e^3)} \,, \\
a_2^\dagger b_1^\dagger  \equiv  E_{-(e^2-e^3)} \,, \quad & a_2 b_1 \equiv  E_{e^2-e^3} \,, \\
a_2^\dagger b_2^\dagger  \equiv  E_{-(e^1-e^3)} \,, \quad & a_2 b_2 \equiv E_{e^1-e^3} \,. 
\end{array} 
\label{abpairs}
\ee

Fundamental spinor IRR of $so(4,2)$ is of dimension four. We may denote the generators of this representation by the $4 \times 4$ matrices  
$\Sigma_{\mu \nu}^+$, whose relation to $\Gamma$-matrices of appropriate signature and dimension can be found in \cite{Govil:2013uta}. As it is well-known, this representation is not unitary but it may be used to induce the UIR that we are seeking for. To do so, we introduce a four component spinor of the form \cite{Sperling:2018xrm,Govil:2013uta,Govil:2014uwa}
\be
\psi = \left ( 
\begin{array}{c}
a_1^\dagger \\
a_2^\dagger\\
b_1 \\
b_2	
\end{array}	
\right ) \,, \quad {\bar \psi} = \psi^\dagger \Gamma^6 = (- a_1, -a_2, b_1^\dagger, b_2^\dagger) \,.
\ee
A Schwinger-type realization of the $so(4,2)$ algebra is then provided by $M_{\mu \nu} = {\bar \psi} \Sigma_{\mu \nu}^+ \psi$ and gives a unitary representation of  $so(4,2) \equiv su(2,2)$ on the Fock space generated the action of $a_\alpha^\dagger$ $b_\alpha^\dagger$ on the vacuum state with unit conformal energy. The latter splits into a direct sum of infinite number of unitary irreducible representations \cite{Govil:2013uta,Govil:2014uwa}. We observe that in the oscillator basis, Cartan generator $H_3$ and a $SO(4,2)$ invariant operator ${\widehat N}$ take the forms \cite{Sperling:2018xrm, Govil:2013uta,Govil:2014uwa}: 
\beqa
H_3 &=& M_{56} =  {\bar \psi} \Sigma_{56} \psi = \frac{1}{2} ({\hat N}_a + {\hat N}_b +2) \,, \nn \\
{\hat N} &:=& {\bar \psi}\psi = - {\hat N}_a + {\hat N}_b -2 \,.
\eeqa
In order to explicitly construct the {\it doubleton} representations, let us consider the states labeled by IRRs of the maximally compact subalgebra $su(2)_L \oplus su(2)_R \oplus u(1)$, in the form $|h_3, J_L, J_R \rangle$ on which any combination of annihilation-creation pairs given in (\ref{abpairs}) naturally acts. Clearly, $|1,0,0 \rangle$ constitutes the vacuum state which is annihilated by all $a_i$ and $b_j$. This vacuum state is clearly specified by $N=-2$ eigenvalue of ${\widehat N}$ and has unit conformal energy, i.e. $h_3 = 1$. Based on this vacuum, we can introduce two representations of $so(4,2)$ with the lowest weight vectors, which are given as
\beqa
|1 + \frac{k}{2}, \frac{k}{2}, 0 \rangle \,, \quad N &=& - k -2 \,, \quad k \in {\mathbb Z} \,, \nn \\
|1 + \frac{k}{2}, 0, \frac{k}{2} \rangle \,, \quad N &=&  k -2 \,, \quad k \in {\mathbb Z} \,.
\label{lws1}
\eeqa
The fact that these are the lowest weight vectors of a representation of $so(4,2)$ is easily observed since all $E_{-(e^i \pm e^j)} \subset {\cal L}^-$ annihilate these states as they are built up from combinations of $a_i b_j$. With the action of the ladder operators $E_{\pm (e^i \pm e^j)} \subset {\cal L}^\pm$ on either of the lowest weights given in (\ref{lws1}), infinite number of states are generated for any given value of $k$. In other words, for each value of $k$, two inequivalent unitary irreducible representations which are infinite-dimensional are generated in this manner. These are called the {\it doubletons} of $so(4,2)$ and the corresponding group $SO(4,2)$. Interchanging $su(2)_L$ and $su(2)_R$ swaps these inequivalent doubletons at a given value of $k$. We can label the {\it doubleton} representations via the eigenvalue $N = \mp k - 2$ of ${\widehat N}$. In order to label all the states in a given {\it doubleton}, in addition to the labels $h_3 ,J_L, J_R$, we also need the eigenvalues of $L_3$ and $R_3$, which we denote as $m_L$ and $m_R$. Thus, we label the states as $|h_3, J_L,J_R, m_L, m_R \rangle$. Roots in ${\cal L}^\pm$ shift $h_3$ to $h_3 \pm 1$ and each of $J_L$, $J_R$ by $ \pm \frac{1}{2}$. This means that the eigenvalue of ${\widehat N}$ is preserved under the action of the roots. In other words, ${\widehat N}$ commutes with all the generators, verifying that it is an invariant operator as previously claimed.  

For the eigenvalues of $H_3$ and ${\widehat N}$, we may write in terms of $J_L$ an $J_R$
\beqa
h_3 &=& 1 +J_L +J_R \,,  \nn \\
N &=& -N_a +N_b-2 = - 2 J_L + 2 J_R -2 = \mp k- 2 \,,
\label{h3n}
\eeqa
where $\mp$ sign in the r.h.s. of the last equality on the second line specifies the two inequivalent {\it doubleton} representation given in (\ref{lws1}). Inverting these equations we have,
\be
J_L = \frac{1}{2} h_3 \pm \frac{1}{4} k - \frac{1}{2} \,, \quad J_R = \frac{1}{2} h_3 \mp \frac{1}{4} k - \frac{1}{2} \,. 
\ee
Let us work with the doubleton representation that corresponds to the upper sign in (\ref{h3n}). Action of the roots on the states $|h_3, J_L,J_R, m_L, m_R \rangle$ have the explicit form  
\beqa
E_{\pm (e^1 + e^2)} |h_3, m_L, m_R \rangle &=&  \sqrt{(\frac{1}{2} h_3 + \frac{1}{4} k  \pm m_L + \frac{1}{2})(\frac{1}{2} h_3 + \frac{1}{4} k \mp m_L - \frac{1}{2})} |h_3, m_L \pm 1, m_R \rangle \,,  \\
E_{\pm (e^1 - e^2)} |h_3,  m_L, m_R \rangle &=&  \sqrt{(\frac{1}{2} h_3 - \frac{1}{4} k  \pm m_R + \frac{1}{2})(\frac{1}{2} h_3 - \frac{1}{4} k \mp m_R - \frac{1}{2})} |h_3,  m_L, m_R \pm 1 \rangle \,,  \nn \\
E_{\pm (e^1 +e^3)}|h_3,  m_L, m_R \rangle &=& \sqrt{(\frac{1}{2} h_3 + \frac{1}{4} k +m_L \pm \frac{1}{2})(\frac{1}{2} h_3 - \frac{1}{4} k + m_R \pm \frac{1}{2})}|h_3 \pm 1, m_L \pm \frac{1}{2}, m_R \pm \frac{1}{2} \rangle \,, \nn \\
E_{\pm (e^2 + e^3)} |h_3, m_L, m_R \rangle &=& \sqrt{(\frac{1}{2} h_3 + \frac{1}{4} k + m_L \pm \frac{1}{2})(\frac{1}{2} h_3 - \frac{1}{4} k - m_R \pm \frac{1}{2})} |h_3 \pm 1, m_L \pm \frac{1}{2}, m_R \mp  \frac{1}{2} \rangle \,, \nn \\
E_{\mp (e^2 - e^3)} |h_3,  m_L, m_R \rangle &=& \sqrt{(\frac{1}{2} h_3 + \frac{1}{4} k - m_L \pm \frac{1}{2})(\frac{1}{2} h_3 -\frac{1}{4} k + m_R\pm \frac{1}{2})}  |h_3 \pm 1, m_L \mp \frac{1}{2}, m_R \pm \frac{1}{2} \rangle \,, \nn \\
E_{\mp (e^1 - e^3)} |h_3, m_L, m_R \rangle &=& \sqrt{(\frac{1}{2} h_3 +\frac{1}{4} k - m_L \pm \frac{1}{2})(\frac{1}{2} h_3 - \frac{1}{4} k - m_R\pm \frac{1}{2})} |h_3 \pm 1, m_L \mp \frac{1}{2}, m_R \mp \frac{1}{2} \rangle \,. \nn
\label{doubletonbasis}
\eeqa
As for the {\it doubleton} representation with the lower sign in (\ref{h3n}), we can simply take $k \rightarrow - k$ in the coefficients provided above\footnote{We could have used a notation with $\pm k $ and $\mp k$ in the expression above to indicate these distinct doubleton representations, but that notation interferes with the $\pm$ and $\mp$'s associated to the root pairs, therefore we avoid the use of such a notation.}. 

We are now in a position to state one of the most crucial result in this article. Namely, we observe that the infinitely degenerate states of the $4D$ model at each energy level on either of the chiral components can be labeled by one or the other of the {\it doubleton} representations of $SO(4,2)$ with $N=-3$ or $N=-1$, since for either of the two we can then match the eigenvalues $h_3 = \frac{3}{2} \,, \frac{5}{2} \,, \cdots$ of $H_3$ with the eigenvalues $l^\prime =  \frac{3}{2} \,, \frac{5}{2} \,, \cdots $ of the operator $A$. In other words, spectrum of $A$ matches in a one to one and onto manner with that of $H_3$ on the states spanning these {\it doubleton} representations. Hence, from now on we make the identification $|h_3, m_L, m_R \rangle \equiv |\ell^\prime, m_L, m_R \rangle$ for the UIRs with $N=-3$ or $N=-1$. Comparing (\ref{extendedcom1}) with (\ref{comso42part1}) and (\ref{comso42part2}) we further infer the identifications
\beqa
&& \frac{1}{\sqrt{S}} K^1_+ \equiv  E_{e^1 +e^3} \,, \quad \frac{1}{\sqrt{S}} K^2_+ \equiv   E_{-(e^1 - e^3)}  \,, \quad \frac{1}{\sqrt{S}} K^3_+ \equiv   E_{e^2 +e^3}  \,, \quad \frac{1}{\sqrt{S}} K^4_+ \equiv  E_{-(e^2 -e^3)} \,, \nn \\
&& K^1_- \frac{1}{\sqrt{S}} \equiv  E_{-(e^1 +e^3)} \,, \quad K^2_- \frac{1}{\sqrt{S}} \equiv   E_{(e^1 - e^3)}  \,, \quad K^3_-\frac{1}{\sqrt{S}} \equiv  E_{-(e^2 +e^3)}  \,, \quad K^4_- \frac{1}{\sqrt{S}} \equiv   E_{(e^2 -e^3)} \nn \,.
\label{KvE}
\eeqa
Acting on the states $|n, l^\prime, m_L, m_R \rangle$, $K_\pm^i$ pick an additional factor of $2 \sqrt{n + l^\prime \pm \frac{1}{2}}$ for $\ell^\prime > 0$ as can be seen using (\ref{KKcom1}), (\ref{KvE}) and (\ref{doubletonbasis}). Concretely, we have
\beqa
L_\pm |n, l^\prime, m_L, m_R \rangle &=&  \frac{1}{2} \sqrt{(\ell^\prime  \pm 2 m_L + \frac{3}{2})( l^\prime \mp 2 m_L - \frac{1}{2})} |n, l^\prime, m_L \pm 1, m_R \rangle \,,  \\
R_\pm |n, l^\prime, m_L, m_R \rangle &=&  \frac{1}{2} \sqrt{(\ell^\prime \pm 2 m_R + \frac{1}{2})(l^\prime  \mp 2 m_R - \frac{3}{2})} |n, l^\prime,  m_L, m_R \pm 1 \rangle \,,  \nn \\
K_\pm^1|n, l^\prime,  m_L, m_R \rangle &=& \sqrt{(l^\prime + \frac{1}{2}  + 2 m_L \pm 1)(l^\prime - \frac{1}{2}  + 2 m_R \pm 1 )(n + l^\prime \pm \frac{1}{2})}  |n, l^\prime \pm 1, m_L \pm \frac{1}{2}, m_R \pm \frac{1}{2} \rangle \,, \nn \\
K_\pm^2 |n, l^\prime, m_L, m_R \rangle &=&  \sqrt{(l^\prime +\frac{1}{2}  - 2 m_L \pm 1)(l^\prime - \frac{1}{2}  - 2 m_R\pm 1)(n + l^\prime \pm \frac{1}{2})} |n, l^\prime \pm 1, m_L \mp \frac{1}{2}, m_R \mp \frac{1}{2} \rangle \,, \nn \\
K_\pm^3|n, l^\prime, m_L, m_R \rangle &=&   \sqrt{(l^\prime + \frac{1}{2}  + 2 m_L \pm 1)(l^\prime - \frac{1}{2}  - 2 m_R \pm 1)(n + l^\prime \pm \frac{1}{2})} |n, l^\prime \pm 1, m_L \pm \frac{1}{2}, m_R \mp  \frac{1}{2} \rangle \,, \nn \\
K_\pm^4|n, l^\prime,  m_L, m_R \rangle &=& \sqrt{(l^\prime + \frac{1}{2}  - 2 m_L \pm 1)(l^\prime -\frac{1}{2}  + 2 m_R\pm 1)(n + l^\prime \pm \frac{1}{2})}| n, l^\prime \pm 1, m_L \mp \frac{1}{2}, m_R \pm \frac{1}{2} \rangle \,. \nn
\label{Kbasis}
\eeqa
The foregoing discussion makes the identification of the extended symmetry generators with either of the $SO(4,2)$ {\it doubleton} representation with $N=-3$ or $N=-1$ manifest and the either of the representations can be used to enumerate the infinite fold degeneracy of the flat LL spectra of the model given in (\ref{spec2}). 

For $\ell^\prime < 0$, i.e. the negative helicity component of the spectrum, energy levels are only finitely degenerate. We easily see from (\ref{spec2}) that at $E=3$, only possible value of $l^\prime$ is $-\frac{3}{2}$, while for $E=4$, the possible values for $\ell^\prime$ are $-\frac{3}{2}$ and $-\frac{5}{2}$ and in general for $E \geq 3$ the possible values of $\ell^\prime$ are $-\frac{3}{2}, -\frac{5}{2}, \cdots ,(\frac{3}{2} - E)$. To label these degenerate states, we essentially need the representations defined through their highest weight states, i.e. the complex conjugate representation. The latter can be obtained from the {\it doubletons} defined via (\ref{lws1}) by taking $(h_3, k) \rightarrow (- h_3 ,-k)$ and making the exchange\footnote{To be more precise, these representations are generated by $-M_{ab}^*$.} $J_L \leftrightarrow J_R$. Complex conjugate {\it doubleton} representations are bounded from above and those with $N = -3^*$ and $N=-1^*$ are the two inequivalent UIRs that may be used. The physical operators $K_{\pm}^i$ acting on these complex conjugate representations bring a factor $\sqrt{E + l^\prime -1 \pm \frac{1}{2}}$, where $E$ stands for the energy eigenvalue\footnote{Note that, in terms of the energy eigenvalues, the factor $\sqrt{(n + l^\prime \pm \frac{1}{2})}$ that appears in (\ref{Kbasis}) also takes the form $\sqrt{E + l^\prime -1 \pm \frac{1}{2}}$. We already know that, $K_\pm^i$ are commuting with the Hamiltonian, however the negative helicity part of the spectrum is not independent of $l^\prime$, therefore it is imperative to express this factor in terms of the energy to make the proper physical interpretation manifest, while it makes no difference to write it in terms of $n$ or $E$ for the positive helicity part as $E$ does not depend on $l^\prime$.}. We see that the $K_-^i$ annihilate the states with $l^\prime \leq \frac{3}{2}-E$, with $E = 3,4,\cdots$, fitting perfectly with the observed finitely degenerate spectrum.  

Let us also recall that the spin operator $S_{ab}$ have the chiral components $S_{ab}^+$ and $S_{ab}^-$ and the spectrum (\ref{spec2}) is the same in each chiral branch.
These chiral parts are mapped to each other upon interchanging the left- and the right- generators of $SU(2)_L \times SU(2)_R$. From these facts, we immediately infer that, we can employ both of the {\it doubleton} representations with $N=-3$ and $N=-1$ one enumerating the infinite degeneracy in the left-chiral and the other in the right-chiral component for the flat spectra with positive helicity. Similarly both of the complex conjugate UIRs with $N = -3^*$ and $N=-1^*$ can be employed to label the negative helicity part of the spectrum. In this manner all the degeneracies in the spectrum (\ref{spec2}) are accounted for.     

\section{Generalization to $d$-dimensions}

It is essentially rather straightforward to generalize the $4D$ model and the preceding developments to $d$-dimensions. With the  
$SO(d)$ gauge field $G = 2 m \omega r_b S_{ab}$, (\ref{H1}) generalizes to
\be
H_d = \frac{1}{2m} (p_a - G_a)^2 -  \frac{d-2}{2} m \omega^2 r_a^2 \,,
\label{Hd1}
\ee
while (\ref{H2}) has the same formal structure
\be
H_d = \frac{p_a^2}{2m} + \frac{1}{2} m \omega^2 r_a^2 - \omega \sum_{a <b = 1 }^{d}  L_{ab} \Gamma_{ab} \,,
\label{Hd2}
\ee
where now $L_{ab} : = r_a p_b - r_b p_a \,, (a,b= 1,\cdots,d)$ are the orbital angular momentum operators that span the $(l,0,\cdots,0)$ IRR of $SO(d)$, which is of dimension ${\cal N} = (d + 2 l -2) \frac{(d+l+3)!}{l! (d-2)!}$. The Casimir operator in this IRR of $SO(d)$ satisfies $\sum_{a<b}L_{ab}^2 = l (l + d - 2) \mathds{1}_{{\cal N}}$. $\Gamma_{ab}$ are proportional to the spin operator $S_{ab}$ in $d$-dimensions and can be given in terms of the commutators of the $\Gamma$-matrices in $d$-dimensions as $S_{ab} := \frac{1}{2} \Gamma_{ab} : = -\frac{i}{4} \lbrack \Gamma_a \,, \Gamma_b \rbrack$. For $d$ odd, $d= 2k+1$, $\Gamma_a$ are of rank $k$; they are $2^k \times 2^k$ matrices and there are $2k+1$ of them. $SO(2k+1)$ has rank $k$ and $S_{ab}$ spans the fundamental spinor representation $(\frac{1}{2},\frac{1}{2}\,,\cdots \,, \frac{1}{2})$ of $SO(2k+1)$ which is $2^k \times 2^k$-dimensional. For $d$ even, $d=2k+2$, $\Gamma_a$ has rank $k$, $2^{k+1} \times 2^{k+1}$ matrices and $S_{ab}$ span a reducible representation of $SO(2k+2)$, which decomposes as $S_{ab} = S_{ab}^+ \oplus S_{ab}^-$ to the fundamental left- and right-chiral spinor representations $(\frac{1}{2},\frac{1}{2}\,,\cdots \,, \pm \frac{1}{2})$, which are each $2^k \times 2^k$-dimensional. The chiral projections to $S_{ab}^\pm$ can be obtained using the projection operators ${\cal P}^\pm = \frac{1}{2} (1 \pm \Gamma_{2k+3})$, where $\Gamma_{2k+3} := \Gamma_1 \Gamma_2 \cdots \Gamma_{2k+2}$. In terms of the $\Gamma$-matrices of rank $k$, we may write $S_{ab}^\pm \equiv (S_{ij}, S_{i 2k+2}) := (S_{ij}, \pm \frac{1}{2} \Gamma_i)$ with $i,j = 1,\cdots, 2k+1$.  The facts listed above are well-known and the spectrum of $H$ is already given in \cite{LiWu, LiWuSup}. For completeness, we provide the essential results here, to lay out the foundations for the developments that will ensue. Eigenvalues of the SO term follows from a similar calculation as in the $4D$ case and they are given as
\beqa
\sum_{a<b} L_{ab} S_{ab}^\pm  &=& 
\begin{dcases}
	\quad \frac{l}{2} & \mbox{on} \left( l +\frac{1}{2} \,,\frac{1}{2}, \cdots \,, (\pm) \frac{1}{2} \right) \,, \quad \mbox{i.e. spin} \, \uparrow \,, \\
	-\frac{l + d -2}{2} & \mbox{on} \left( l - \frac{1}{2} \,, \frac{1}{2} \,,\cdots \,, (\pm) \frac{1}{2} \right) \,, \quad \mbox{i.e. spin} \, \downarrow	
\end{dcases}
\label{SOspecd}	
\,.
\label{SOterm_spec_ddim}
\eeqa
It should be clear that the $(\pm)$ in (\ref{SOspecd}) distinguishes the left- and the right-chiral representations for $SO(2k+2)$; while for $SO(2k+1)$ only the upper sign appears. This gives the spectrum of $H_d$ in (\ref{Hd1}) or equally in (\ref{Hd2}) as 
\be
E =
\begin{dcases}
	 2 \omega \left (n + \frac{d}{4} \right) \,, & \mbox{spin $\uparrow$}	\\
	 2 \omega \left (n + l + \frac{3}{4} d - 1 \right) = 2 \omega \left ( n - l^\prime + \frac{d}{4} + \frac{1}{2} \right) \,, & \mbox{spin $\downarrow$}	
	 \label{specd}
\end{dcases} \,.	
\ee	
Note that as in the $4D$ model, the spectrum in the positive helicity branch is flat, indicating an infinite-fold degeneracy for this part of the spectrum. We also stick to the choice of the parameter values $m=1$ and $\omega = \frac{1}{2}$. In analogy with the $3D$ \cite{Haaker} and $4D$ results, we may introduce the operator 
\be
A = \sum_{a < b} L_{ab} \Gamma_{ab} + \frac{d-1}{2} \,,
\label{opAd}
\ee
whose eigenvalues may still be denoted as $l^\prime$. We have $l^\prime = l + \frac{d-1}{2}$ for the positive and $l^\prime = -l - \frac{d-3}{2}$ for the negative helicity components so that $ l^\prime = \pm \frac{d-1}{2} \,, \pm \frac{d+1}{2} \,, \cdots$. The eigenstates of the Hamiltonian $H_d$ can be denoted by the kets $|n, l^\prime\,,[s]_{SO(d)} \,, [m]_{SO(d)} \rangle$, where $[s]_{SO(d)}$ stands as a collective index for the $SO(d)$ UIR and  $[m]_{SO(d)}$ as a collective index of the quantum numbers within this UIR of $SO(d)$ that unambiguously label these eigenstates.

Using the operator $A$, we can introduce the $d$-dimensional vector operators $M_a$ and $N_a$ exactly in the same formal form as given in (\ref{vecopsMN}), except that the indices $a,b$ are now taking values in the interval $(1,\cdots,d)$. Total angular momentum operators $J_{ab}= L_{ab} + S_{ab}$, $A$ and appropriately scaled linear combinations of $M_a$, $N_a$ span $\frac{1}{2}(d+2)(d+1)$-dimensional group, which can be identified with the non-compact group $SO(d,2)$. 
%on the eigenstates of the $d$-dimensional SHO with the critical SO term after suitable scalings of the operators $M_a$ and $N_a$ are performed.
This essentially works in the same manner, as we have laid out in detail for the $4D$ case. We may use the discrete unitary irreducible representations of $SO(d,2)$ defined through their lowest weight vectors to label the infinite degeneracy of the flat part of the spectrum in (\ref{specd}). $so(d,2)$ is of rank $k+1$ for $d=2k+1$ and of rank $k+2$ for $d = 2k+2$. The relevant discrete UIR of $so(d,2)$ can be constructed using the three-graded decomposition of the Lie algebra $so(d,2)$ w.r.t its maximally compact subalgebra \cite{Fernando:2015tiu, Gunaydin:2016bqx} 
\be
so(d,2) \equiv {\cal L}^+ \oplus {\cal L}^0 \oplus {\cal L}^- \,,
\label{threegrading2}
\ee
where ${\cal L}^0$ stands for the maximally compact subalgebra $so(d) \oplus u(1)$. ${\cal L}^\pm$ which contains the remaining generators of $so(d,2)$, and the three-grading has the same structure as defined before in (\ref{threegrading1}). From the existing literature, it is readily known that the discrete unitary representations of $so(d,2)$ can be labeled by the eigenvalues of the $U(1)$- generator and these representations generalize the {\it singleton} representation of $so(3,2)$ for odd values of $d$ and the {\it doubleton} representations of $so(4,2)$ for even values of $d$ \cite{Fernando:2015tiu, Gunaydin:2016bqx}. For $d=2k+2$, the representations we need fall into the class in which the $SO(2k+2)$ subgroup carry the IRRs $(\frac{s}{2}, \frac{s}{2},\cdots\, \pm \frac{s}{2}) \equiv [s]_{SO(d)}$, where $s$ is a non-negative integer. Corresponding to each of these representations there is a UIR of $so(d,2)$ with the lowest weight vectors $|\frac{1}{2} (d+s-2)\,, (\frac{s}{2}, \frac{s}{2},\cdots\, \pm \frac{s}{2}) \rangle$, whose $U(1)$-charge, i.e. the eigenvalue of the $(k+2)^{th}$ Cartan generator $H_{k+2}$ is $\frac{1}{2} (d+s-2)$. Action of the operators in the ${\cal L}^-$ sector of the three-grading annihilate these lowest weight states, while the repeated action of operators ${\cal L}^+$ generates these UIRs. In particular, action of ${\cal L}^\pm$ shifts the eigenvalue of $H_{k+2}$ by $\pm 1$, and map $[s]_{SO(d)}$ to $[s\pm 1]_{SO(d)}$, while also changing the collective $[m]_{SO(d)}$ indices; ${\cal L}^\pm$ can be spanned by the roots of $so(d,2)$ in the Cartan-Weyl basis, whose organization is determined by the fact that roots in ${\cal L}^\pm$ should shift the eigenvalue of $H_{k+2}$ by $\pm 1$. 

Among these UIRs, we need the one with $s=1$, whose $U(1)$ charge in the lowest weight sector matches with the lowest possible positive eigenvalue $\frac{d-1}{2}$ of $A$. Thus, the spectrum of $A$ matches with that of $H_{k+2}$ on these UIRs and the states in either of them with the lowest weights $|\frac{1}{2} (d-1)\,, (\frac{1}{2}, \frac{1}{2},\cdots\, \pm \frac{1}{2}) \rangle$ span the infinite degeneracy of the flat spectrum in (\ref{specd}). There are overall $2d$ different linear combinations $K_\pm^i$ ($i:1,\cdots,d$) of $M_a$'s and $N_a$'s such that $\frac{1}{\sqrt{S}} K_+^i$ span ${\cal L}^+$ and $ K_-^i \frac{1}{\sqrt{S}}$ span ${\cal L}^-$, where $S = H + A - \frac{1}{2} \left( \frac{d}{2} +1 \right)$ up to an overall constant which is immaterial for our present purposes. Just like the $4D$ case, we can associate one of these UIRs with the left- and the other with the right- chiral component to label and distinguish the degenerate spectrum. For the negative helicity states, energy disperses with the eigenvalues of $A$ and the degeneracy is finite. The complex conjugate representations, which are practically obtained by $h_{k+2} \rightarrow - h_{k+2}$, have the highest weight vectors $|- \frac{1}{2} (d-1)\,, (\frac{1}{2}, \frac{1}{2},\cdots\, \pm \frac{1}{2}) \rangle$ and can be used to label the degenerate states in this branch, noting that the unscaled operators $K_{-}^i$ annihilate the states with $l^\prime < \frac{d}{4} + \frac{1}{2} - E$, with $E$ taking on the values $ \frac{3}{4} d, \frac{3}{4} d + 1, \cdots$. For $d=4$, our previously determined result is immediately obtained, while, for instance, for $d=6$, negative helicity states have the lowest energy $\frac{9}{2}$ and therfore no states with $l^\prime < - \frac{5}{2}$ exist, in perfect agreement with the observed spectrum and degeneracy of the negative helicity states.

For odd values of $d$, the relevant representation of $so(d,2)$ is also labeled by the $U(1)$ charge, and has the value $\frac{d-1}{2}$ for the lowest weight state \cite{Fernando:2015tiu, Gunaydin:2016bqx}. The latter is given by $|\frac{1}{2} (d-1)\,, (\frac{1}{2}, \frac{1}{2},\cdots\, \frac{1}{2}) \rangle$, where the $(\frac{1}{2}, \frac{1}{2},\cdots\, \frac{1}{2})$ is the $2^{(\frac{d-1}{2})}$-dimensional fundamental spinor IRR of $SO(d)$. These states are annihilated by all the operators belonging to ${\cal L}^-$ of the three grading and a UIR of $so(d,2)$ is generated by the repeated application of the operators in ${\cal L}^+$. It can be readily noted that, as opposed to the infinite family of representations for even $d$ (corresponding to the pair of UIRs labeled by the integer $s$), for $d$ odd there is only a unique spinoral UIR of $so(d,2)$. For $d=3$, this is nothing but the Dirac {\it singleton} representation with spin $\frac{1}{2}$. In this UIR spectrum of $A$ identifies with that of the $U(1)$ generator, which may be taken as the $(k+1)^{th}$ Cartan generator $H_{k+1}$, and the states generated from the lowest weight $|\frac{1}{2} (d-1)\,, (\frac{1}{2}, \frac{1}{2},\cdots\, \frac{1}{2}) \rangle$ completely label the infinite degeneracy of the flat spectrum. The rest of the correspondence is almost the same as that of even $d$ given in the previous paragraph, except that for $d$ odd, the operator $K_\pm^{2k+1}$ associated to the last root pair $E_{\pm e^k}$, in the form $E_{e^k} \equiv \frac{1}{\sqrt{S}} K_+^{2k+1}$ and $E_{-e^k} \equiv K_i^{2k+1} \frac{1}{\sqrt{S}}$ are obtained from the linear combinations of only the $(2k+1)^{th}$ components of $M_a$ and $N_a$, i.e. $K_\pm^{2k+1} = \frac{1}{\sqrt{2}} (M_{2k+1} \mp i N_{2k+1})$, while the other $K_\pm^i$ are linear combinations that involve two components from each of $M_a$ and $N_a$ (\ref{KMN}). Thus, all $K_\pm^{i}$ shift the eigenvalue of $H_{k+1}$ by $\pm 1$, and $[s]_{SO(d)}$ to $[s\pm 1]_{SO(d)}$, while $[m]_{SO(d)}$ change accordingly under $K_\pm^{-}$, $(i \neq 2k+1)$, but remains unchanged under $K_\pm^{2k+1}$. For $d=3$, this result can be seen from the formula provided in \cite{Haaker} and is a characteristic discriminating odd $d$ from even $d$.  

\section{Spectrum Generating Algebra}

Using the annihilation and creation operators 
\be
c_a = \frac{1}{2} r_a - i p_a \,, \quad c_a^\dagger = \frac{1}{2} r_a + i p_a \,, \quad \lbrack c_a \,, c_b^\dagger \rbrack = \delta_{ab} \,, \quad (a = 1\,,\cdots \,, d )\,,
\ee
we may express the Hamitonian $H_d$ in (\ref{Hd2}) as
\be
H = \frac{1}{2} \left ({\widehat N}_c +  \frac{d}{2} \right) - \frac{1}{2} \sum_{a<b} L_{ab} \Gamma_{ab} \,,
\ee
where ${\widehat N}=c_a^\dagger c_a$ and the orbital angular momentum operator can be expressed in terms of $c_a \,, c_a^\dagger$ as $L_{ab} = - i (c_a^\dagger c_b - c_b^\dagger c_a )$.

Generalizing the discussion given in \cite{Haaker}, we may introduce the operators ${\cal S}_\pm$, which are quadratic in $c_a$'s and $c_a^\dagger$'s as 
\be
{\cal S}_+ = - \frac{1}{2} c_a^\dagger c_a^\dagger \,, \quad {\cal S}_- = -\frac{1}{2} c_a c_a \,.
\ee
Since, $\lbrack {\hat N} \,, {\cal S}_\pm \rbrack = \pm 2 {\cal S}_\pm$ and $\lbrack L_{ab} \,, {\cal S}_\pm \rbrack = 0$, we infer that ${\cal S}_\pm$ shifts the energy eigenvalues by $\pm 1$. It is straightforward to show that
\be
\lbrack H \,, {\cal S}_\pm \rbrack = \pm {\cal S}_\pm \,, \quad \lbrack {\cal S}_+ \,,{\cal S}_- \rbrack = - 2 \left( H + \frac{A}{2} - \frac{d-1}{4} \right) \,.
\label{so21com}
\ee
Upon identification of $ H + \frac{A}{2} - \frac{d-1}{4}$ with the ${\cal S}_3$ generator, (\ref{so21com}) corresponds to the $SO(2,1)$ commutation relations. In particular, ${\cal S}_3 \,, {\cal S}_\pm$ span the unitary irreducible representation of $SO(2,1)$ with the extremal weights $\Lambda = \frac{l^\prime}{2} + \frac{1}{4}$ for $\ell^\prime = \frac{d-1}{2}\,, \frac{d+1}{2} \,, \cdots$ and $\Lambda = - \frac{l^\prime}{2} + \frac{3}{4}$ for $l^\prime = - \frac{d-1}{2}\,, -\frac{d+1}{2} \,, \cdots$. In these representations of $SO(2,1)$, eigenvalues of the Casimir operator $\frac{1}{2}({\cal S}_+ {\cal S}_- + {\cal S}_-{\cal S}_+)- {\cal S}_3^2$ are given as $\Lambda (1 - \Lambda)$. Using this information and (\ref{so21com}) 
we easily find 
\beqa
{\cal S}_\pm | n \,, l^\prime, m_L \,, m_R \rangle &=& \sqrt{\left (n + \frac{1}{2} \pm \frac{1}{2}\right) \left(n +  l^\prime \pm \frac{1}{2} \right )}  | n \pm 1 \,, l^\prime, m_L \,, m_R \rangle \,, \quad l^\prime > 0 \,,  \\
{\cal S}_\pm | n \,, l^\prime, m_L \,, m_R \rangle &=& \sqrt{\left (n + \frac{1}{2} \pm \frac{1}{2}\right) \left(n -  l^\prime + 1 \pm \frac{1}{2} \right )}  | n \pm 1 \,, l^\prime, m_L \,, m_R \rangle \,, l^\prime < 0 \\
&\equiv& \sqrt{\left (E + l^\prime - \frac{d}{4} \pm \frac{1}{2}\right) \left(E - \frac{d}{4} + \frac{1}{2} \pm \frac{1}{2} \right )}  | E \pm 1 \,, l^\prime, m_L \,, m_R \rangle \,. \nn
\label{so21states}
\eeqa
General considerations on the UIRs of $SO(2,1)$ require that $\Lambda \geq \frac{1}{2}$ \cite{Perelomov}, and this is fulfilled in the present case since $|l^\prime| \geq \frac{d-1}{2}$. For both the positive and negative helicity components the lowest weight state is $| 0 \,, l^\prime, m_L \,, m_R \rangle$. Since the energy spectrum is $E = n - l^\prime + \frac{d}{4} + \frac{1}{2}$ for negative helicity states, it is readily inferred from the second line of (\ref{so21states}) that ${\cal S}_-$ annihilates the states with $l^\prime < \frac{d}{4} + \frac{1}{2} - E$ in accord with the result determined in the preceding section.

\section{Discussion and Conclusions}

In this paper we have examined the degeneracies in the energy spectrum of $d \geq 4$-dimensional SHOs coupled to Aharanov-Casher type $SO(d)$ gauge fields. The Hamiltonians of these models can equally be expressed as SHOs, with a spin-orbit terms, whose coupling strength is tuned to the SHO frequency. With our choice of sign for the SO coupling the positive helicity part of the energy spectrum is flat and led to the interpretation of these models as TR invariant LLs in higher dimensions \cite{LiWu}, generalizing the QSHE \cite{Bernevig:2006zz}. Focusing on the $4D$ model, we have introduced two vector operators commuting the Hamiltonian and succeeded in demonstrating that the symmetry group $SO(4)$ of the model extends to the non-compact dynamical symmetry $SO(4,2)$ and that the discrete UIRs of this group, the so called {\it doubletons} specified via the invariants $N=-3$ and $N=-1$ provide the complete labeling of the infinite degeneracy of the flat spectrum. Subsequently, all of these results generalized to the models in $d$-dimensions and shown that the extended non-compact symmetry group is indeed $SO(d,2)$ and the infinite degeneracy of the flat spectra is completely accounted for by exploiting the discrete series UIRs of $SO(d,2)$, which generalize the {\it singleton} and the ${\it doubleton}$ representations in $d=3$ and $d=4$ to all odd and even dimensions, respectively. 

Since $SO(d,2)$ is the isometry group of $AdS_{d+1}$, we may contemplate that the LL states are essentially organized  w.r.t. an underlying radially ``deformed" AdS geometry rather than the flat space, due to the extended non-linear symmetry generated by the operators $J_{ab}$, $A$ and $K_\pm^i$, whose commutation relations involve non-linear terms as manifestly seen from (\ref{comMN}) and (\ref{KKcom1}). The spectrum of the operator $A$ in (\ref{opAd}) matches with that of the $U(1)$ generator in the UIR of $SO(d,2)$ in which the latter has the eigenvalue $\frac{d-1}{2}$ in the lowest weight states. Considering the $SO(d,2)$ as either the conformal group for $d$-dimensional Minkowski space-time or, as above, the isometry group of $AdS_{d+1}$, spectrum of the $U(1)$ generator corresponds to the conformal energy or the AdS energy \cite{Fernando:2015tiu, Gunaydin:2016bqx}. Under the action of $so(d,2)$ ladder operators, eigenvalues of $A$ shift by $\pm 1$ within the degenerate states, giving further evidence toward the aforementioned interpretation.

Finally, let us note that imposing open boundary conditions say at radius $R_0$, the flat spectrum is no longer maintained once the angular momentum value exceeds a critical value, which depends on the LL and can be numerically estimated for a given model, as it was done for $3D$ case in \cite{LiWu, LiWuSup}. Starting around this critical value, the energy spectrum becomes dispersive indicating the emergence of states localized on the boundary \cite{LiWu}. In fact, the energy spectrum at the surface can be linearized around the Fermi angular momentum and becomes essentially governed by the Hamiltonian 
\beqa
H_{surface} &=& \frac{v_F}{R_0} \sum_{a<b} L_{ab} \Gamma_{ab} - \mu  \nn \\
&=& \frac{v_F}{R_0} \left ( A - \frac{d-1}{2} \right) - \mu \,,
\label{Hsurface}
\eeqa
where $\mu$ stands for the chemical potential. Although $SO(d,2)$ can no longer be considered as the precise extended dynamical symmetry in the presence of the boundary, (\ref{Hsurface}) shifts by $\pm \frac{v_F}{R_0}$ under the action of $so(d,2)$ ladder operators, which suggests the interpretation of $SO(d,2)$ as an effective spectrum generating algebra for the surface states. We think that these brief remarks merit further study and any future progress on them will be reported elsewhere.

 \vskip 1em

\noindent{\bf \large Acknowledgments}

\vskip 1em

\noindent Part of S.K.'s work was carried out during his sabbatical stay at the physics department of CCNY of CUNY and he thanks V.P. Nair and D. Karabali for the warm hospitality at CCNY and the metropolitan area. S.K. acknowledges the financial support of the Turkish Fulbright Commission under the visiting scholar program and the METU research project GAP-105-2018-2809. G.Ü acknowledges the support of the TUBİTAK 2218 post doctoral scholarship program.

\end{document}